**Antineutrino Detectors Remain Impractical for Nuclear Explosion Monitoring**

Michael Foxe[1], Theodore Bowyer[1], Rachel Carr[2], John Orrell[1], Brent VanDevender[1]

1) Pacific Northwest National Laboratory, Richland, WA, USA 99352
2) Massachusetts Institute of Technology, Cambridge, MA, USA 02139

Michael.Foxe@pnnl.gov

**Abstract:** Fission explosions produce large numbers of antineutrinos. It is occasionally asked whether this distinctive, unshieldable emission could help reveal clandestine nuclear weapon explosions. The practical challenge encountered is that detectors large enough for this application are cost prohibitive, likely on the multi-billion-dollar scale. In this paper, we review several hypothetical use cases for antineutrino detectors as supplements to the seismic, infrasound, hydroacoustic, and airborne radionuclide sensors of the Comprehensive Nuclear-Test-Ban Treaty Organization's International Monitoring System. In each case, if an anti-neutrino detector could be constructed that would compete with existing capabilities, we conclude that the cost would considerably outstrip the value it might add to the existing monitoring network, compared to the significantly lower costs for the same or superior capability.

# I. Introduction

The International Monitoring System (IMS) is a network of detectors (seismic, hydroacoustic, infrasound, and radionuclide (particulate and noble gas) throughout the world for the purpose of verification of the Comprehensive Nuclear-Test-Ban Treaty (CTBT). Within the CTBT framework, all nuclear explosion testing is banned, regardless of the location or size of the nuclear explosive test. Waveform technologies (seismic, hydroacoustic, and infrasound) observe a signal that transports at the speed of sound. Airborne radionuclide signals can be observed on a much longer time scale. Observable airborne radionuclide signals may be delayed both due to the subsurface gas migration and atmospheric transport but can be a clear sign of a nuclear explosion (with the ability to confirm the nuclear nature of an explosion). These measurement techniques currently form the technical basis for world-wide monitoring under the CTBT's verification regime.

The combination of these detector technologies offers potential to both estimate the magnitude of events, with a goal of verifying whether a suspicious event is nuclear in nature. Yet, even with the complementary abilities of these technologies, there are scenarios where the detection mechanisms may not yet achieve the desired level of sensitivity achieved by the current IMS. In such situations, it might be advantageous to have a complementary detection technology capable of filling in the measurement gaps or otherwise augment existing IMS detection capabilities. This complementary technology could provide a redundant detection capability for an improved confidence of detection. That motivates the search for new and compelling methods for detection of nuclear explosions.

Since at least the 1950s, physicists have recognized that fission explosions produce intense bursts of antineutrinos primarily from the secondary beta-decay of fission products that occur within seconds of the primary fission event. The scientists who made the first-ever antineutrino detection considered using a nuclear weapon test as the source (Reines, 1996). They would have needed to place their cubic-meter-scale detector within a kilometer of the explosion, due to antineutrino's very weak interaction cross-section. In a kiloton nuclear explosion, the initial fissions produce a variety of fission products resulting in many antineutrinos emitted during the life of those fission products. Of the $10^{24}$ antineutrinos emitted from a kiloton explosion, only a few would have registered in the group's detector, even at a distance of just a few hundred meters. The group found it more practical to deploy their detector outside a nuclear reactor.

Nonetheless, as scientists have built ever-larger neutrino and antineutrino detectors for basic science, the idea of detecting antineutrinos from fission explosions has drawn attention again. After the negotiation of the CTBT, a group at Sandia National Laboratory studied whether antineutrino detectors could play a useful role in verifying the ban. Their study concluded that, "while antineutrino detectors are in theory very attractive for CTBT verification, both engineering difficulties and physics limitations severely limit the actual application of the technology for this purpose" (Bernstein et al*.,* 2001). A more recent study, considering the combination of antineutrino detectors and seismic sensors, came to the same conclusion (Carr et al., 2018). The large size and cost of detectors, as demanded by the low antineutrino interaction probability, are the essential constraints. There was previously a study looking at the cost of a large antineutrino detector network (Guillian, 2006), which found the total cost to be trillions of dollars with the detector technologies at the time.

While nuclear reactors and nuclear explosions both produce antineutrinos, this paper does not focus on the application of antineutrino detectors for nuclear reactor monitoring. Nevertheless, much of the antineutrino detection technology developed within the last two decades for reactor antineutrino monitoring (Bernstein et al., 2010; Bernstein et al., 2001) is relevant as a technology basis for evaluation.

In the following sections, we describe the basic physics behind these studies, and we extend their analyses by giving explicit cost estimates for the antineutrino detectors in three hypothetical use cases: 1) augment the global monitoring network: both confirming the nuclear nature of an explosion observed by seismic sensors and detecting an unobserved nuclear explosion; 2) a co-operative test site transparency scenario; and 3) providing additional information beyond explosion detection.

**II. Physics of Detecting Antineutrinos from a Fission Explosion**

    A. Antineutrino signal from a fission explosion

Antineutrinos are emitted in large quantities from operating nuclear power reactor cores over an extended duration. A nuclear explosion is a rather similar process over a very short time scale (and with less fissile material), resulting in antineutrinos being primarily emitted in a much shorter duration. Thus, a *burst* of antineutrinos would be a distinct signature of an explosive nuclear event. As antineutrinos are very penetrating (due to their low interaction cross section), they are virtually impossible to shield against and may allow for sensitive detectors to make a measurement at some distance. After a fission occurs, the fission fragments undergo a series of beta decays, resulting in five to six antineutrinos per fission that occurs.

For 1-kiloton (kT) nuclear explosions (based on a nuclear explosion detection threshold of ~0.2–0.5 kT (National Academy of Sciences, 2002) – Section III), approximately $10^{24}$ antineutrinos are emitted from the subsequent beta decays. The energy of the emitted antineutrinos ranges from approximately 0 to 10 MeV, with a larger fraction emitted at the lower energies. Of these approximately 30% (2.6 x $10^{23}$ antineutrinos) are emitted in a 10 second window. The other 70% are spread out much later in time and not well time-correlated to the initial fission. For comparison, a 3-GWth nuclear reactor produces approximately $6 \times 10^{21}$ antineutrinos every 10 seconds. The antineutrino flux available for measurement from both of these sources is reduced by the requirement that the energy be above 1.802 MeV for the inverse beta decay reaction ($\bar{v} + p \rightarrow n + e^-$), which is the most prevalent and demonstrated mechanism for measuring reactor antineutrinos. Thus, for a nuclear explosion, the 30% emitted within 10 seconds is reduced by approximately 30% for a total source of detectable antineutrinos (above the 1.802 MeV threshold) of approximately 10% the total (7.25 x $10^{22}$). The flux available at the detector will decrease as the distance between the nuclear explosion and the detector increases. We will investigate the detector size requirements with three subsequent use cases and stand-off-distances.

B. Detection approach

Due to the ability of neutrinos (and antineutrinos) to travel long distances without interacting, very large detectors are required for antineutrino detection. One such detector is the Super Kamiokande (neutrino) detector (Super-K), Figure 1, which consists of 50,000 tons of water. Kamiokande, the predecessor of Super-K, was one of the detectors to observe a time-correlated cluster of neutrinos from Supernova 1987a, that took place approximately 168,000 light-years from Earth (Antonioli, 2004; Hirata, 1987). The correlation of events that are clustered in time can also be extended to the case of a burst of antineutrinos resulting from a nuclear explosion with the same time scale. While Super Kamiokande was built in 1996, Hyper Kamiokande (Hyper-K), with a 260,000-ton water target, is the largest currently proposed antineutrino detector at a cost of ~$1B.

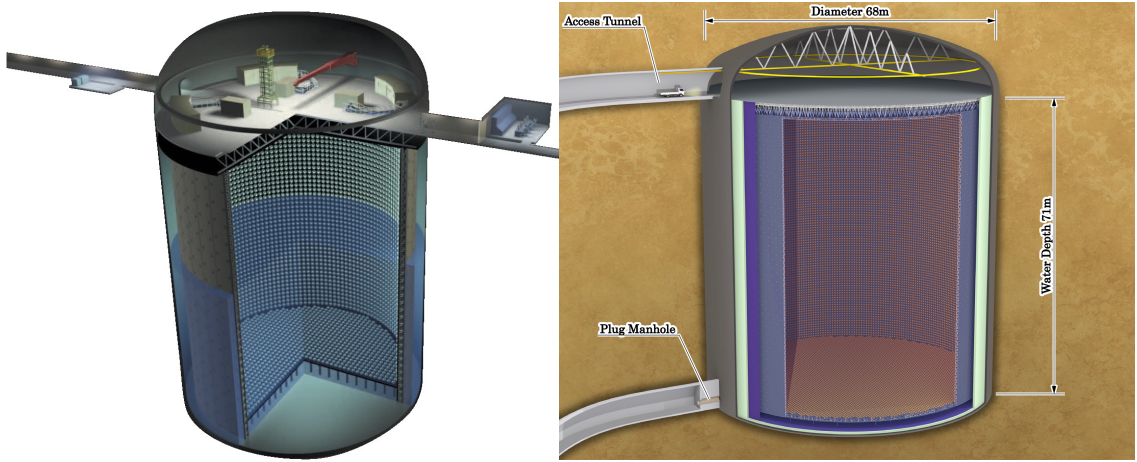

**Figure 1**. Left) Model of the Super Kamiokande detector (Kajita, 2016), demonstrating the size of current neutrino/antineutrino detectors. The total cost of this detector was approximately 100 million dollars. Right) Model of the next generation detector, Hyper Kamiokande (Lodovico, 2017; Hyper-K 2018), expected to cost approximately 1 billion dollars.

More recently, antineutrino detectors have been used to probe the properties of the neutrino/antineutrinos through detection of antineutrinos emitted from a nuclear reactor at distances up to 1 km away (Daya Bay, 2012; Double Chooz, 2012; Reno, 2012). There have also been efforts proposed at monitoring a nuclear reactor at a farther distance (25 km) with a Gadolinium-Doped Water-Cherenkov detector (Askins, 2015). Another smaller antineutrino detector (1 tonne) has previously been placed at the San Onofre Nuclear Generating Station (25 m from the reactor core) to track the antineutrino rate and monitor the amount of uranium and plutonium within the reactor (Bowden et al., 2009). There are now also efforts at monitoring smaller reactors with surface detectors (Ashenfelter, 2018; Huber, 2019)

The simplest detection mechanism for antineutrinos from fission is inverse beta decay. An antineutrino with a minimum energy of 1.802 MeV interacts with a proton creating a positron and neutron. The neutron can subsequently be detected via neutron capture, which is often aided by doping the water with gadolinium. The use of gadolinium decreases the neutron capture time and allows for particle identification to be performed based on the observation of the positron and neutron signals in coincidence (Kim, 2016). One may explore the option of using alternative detection mechanisms (such as neutrino-electron elastic scattering or coherent elastic neutrino-nucleus scattering) to detect the antineutrinos (Hagmann & Bernstein, 2004; Scholberg, 2015), but these other detection mechanisms do not offer the signal detection (Sangiorgio et al., 2013) or discrimination (antineutrino events and background) capabilities required for nuclear explosion monitoring (Foxe et al., 2015).

Throughout the rest of this paper, we assume that inverse beta decay is used as the detection mechanism, which means neutrino backgrounds are not an issue. For this reason, we focus solely on backgrounds that have the potential to mimic antineutrino signals in an inverse beta-decay based detector for nuclear explosion monitoring.

C. Backgrounds to detection

In order to successfully detect antineutrinos from a nuclear explosion, it is important to measure the background sufficiently well to characterize it. These backgrounds consist of: 1) terrestrial antineutrinos, 2) anthropogenic antineutrinos (nuclear reactor antineutrinos), and 3) false antineutrino events (accidental coincidences or other neutron generating primary interactions such as cosmic ray interactions). While accounting for these backgrounds may be difficult, the best remedy is to attempt to reduce the impact of the backgrounds as much as possible. We will assume a best-case scenario for the antineutrino detector: All other potential backgrounds such as noise, instrumental imperfections are eliminated (somehow) resulting in zero extraneous backgrounds.

*False Antineutrino Event Backgrounds:*

Due to the scarcity of detection of antineutrinos, it is important to have sufficient shielding to reduce environmental backgrounds that may mask the antineutrino signal. Such backgrounds include muons and fast neutrons. Since inverse beta decay works on the principle of the detection of the positron followed by the detection of the neutron capture, it is possible for other events (such as fast neutrons) to mimic this behavior. It may be possible to use event location in small detectors to distinguish events from a muon or fast neutron from that of an antineutrino (Sweany et al., 2015). The primary way to reduce the false antineutrino backgrounds is through additional shielding of muons and fast neutrons. The muons and fast neutrons are the results of the cosmic rays that bombard the earth. The best way to reduce the number of muons and fast neutrons is with a large overburden (Mei and Hime, 2006), a feat often performed by deploying a detector deep underground (such as SNOLAB or Gran Sasso (Jillings, 2009)). In these situations, thicker/denser rock between the detector and the surface is preferred as it will better attenuate the muons and fast neutrons from the surface.

While there have been instances when an antineutrino detector has been deployed with little or no overburden, these instances have been close to a nuclear reactor (25 meters away) (Bowden et al., 2009; Haghighat et al., 2018). The increased antineutrino flux near the reactor was able to overcome the background of ~100 false antineutrino events per day (Bowden et al., 2009), but the false antineutrino rate of ~100 events is much higher than the real antineutrino rate from a distant nuclear explosion.

*Antineutrino Backgrounds:*

Antineutrino backgrounds must be handled separately from the other backgrounds that can be reduced through shielding. Antineutrinos occur naturally due to the radioactive decay of various materials within the earth (i.e. uranium and thorium decay chains). While the Earth's crust provides a stable source of antineutrinos, the magnitude varies based on the surrounding geology and geophysics of the location, with the average antineutrino flux being approximately $10^{8.6}$ antineutrinos/100 cm$^2$/s. These antineutrinos are produced from various sources, with different energy ranges (e.g. uranium and thorium decay chains versus the decay of potassium-40). Accounting for the geophysics for a detector location is important to be able to accurately determine the antineutrino background rates. This is a significant source of background when dealing with a detector large enough to observe a nuclear explosion, and the specific detector location would need to avoid any antineutrino hot spots.

Reactor antineutrinos are also an irreducible background, emitted in large quantities from nuclear power reactors. A 1-GW electric nuclear reactor emits ~$10^{21}$ antineutrinos per second, or $10^{22}$ antineutrinos over the 10-second nuclear explosion burst period. As the rate of neutrino emission from a 1-kT nuclear explosion ($10^{24}$) is on the order of a potential source from a close nuclear reactor, the impact of current and future nuclear reactors must be considered when choosing detector locations.

Fortunately for the present study, the combination of all antineutrino backgrounds has been summarized in previous work (Usman, 2015). This prior work looked at the global antineutrino rates as a function of energy and geographic location.

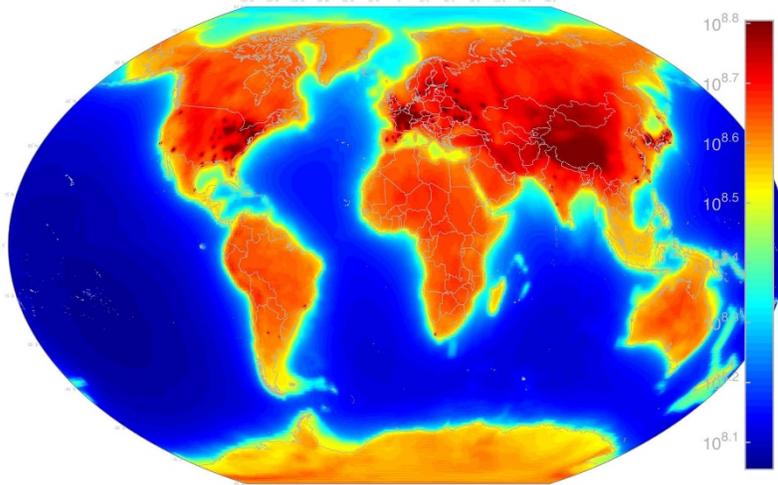

*Figure 2: 2015 survey of all antineutrinos around the world. Approximately 1.6% of the antineutrinos on this map are above 1.8 MeV, with hot spots near nuclear reactors. (Usman, 2015). Flux units are antineutrinos/100 cm$^2$/s at the Earth's surface. Map includes antineutrinos of all energies.*

The study calculated the number of antineutrinos globally based on the sources coming from nuclear reactors and geoneutrinos from the earth's crust (uranium, thorium, and potassium). In the 2015 survey, it was shown that antineutrinos above 1.8 MeV accounted for just 1.6% of the total antineutrino luminosity. Based on this map, the global antineutrino background above the IBD threshold of 1.802 MeV is approximately $10^5$ antineutrinos/cm$^2$/s. This background number is good to keep in mind as it would serve as a detection limit for a land-based antineutrino detector. In order to limit this background and provide flexibility in detector locations, an antineutrino detector deployed in the ocean is a likely scenario.

### III. Hypothetical Uses for Antineutrino Detectors in Explosion Monitoring

Minimum detector size is a parameter that is directly influenced by the antineutrino interaction cross section. Since the antineutrino interaction cross section is very small (~$10^{-44}$ cm$^2$), a detector designed to detect antineutrinos from a source such as a nuclear explosion must include a large number of potential interaction sites for the antineutrino. Additionally, the size further increases as the distance to a nuclear explosion increases and the size of the nuclear explosion decreases.

Based on current nuclear explosion monitoring capabilities, a detection threshold of a nuclear yield of ~0.2–0.5 kT is established (National Academy of Sciences, 2002). For ease of comparison, we choose a 1-kT nuclear explosion as an antineutrino source. Using the source of antineutrinos from a 1-kT nuclear explosion of $7.25 \times 10^{22}$ as calculated in the prior section, we will investigate the size and cost requirements with three subsequent use cases.

**Use Case #1: Support of a Global Monitoring Network**

One way antineutrino detectors could, in principle, supplement the existing IMS is by confirming the nuclear nature of an event picked up by seismic sensors. While seismic waveforms can provide information about the location and yield of a suspected nuclear explosion, they can only indirectly indicate that the event was nuclear in nature. Currently, the IMS uses radionuclide (particulate and noble gas) sensors to seek explicit confirmation that a suspect event involved nuclear fission. Radionuclides are not always released from underground nuclear explosions, and transport to sensor stations is not guaranteed. Antineutrinos are theoretically an attractive alternative to confirm the nuclear nature of an explosion.

*Use Case #1a: Confirming the nuclear nature: seismically cued*

A recent study explored the possibility of using antineutrino detectors to confirm the nuclear nature of suspect seismic events (Carr et al., 2018). The idea is to use the time of detonation inferred from the seismic waveforms as an analysis trigger, or "cue," for an antineutrino detector. One would look for an antineutrino signal in a short window following the seismic cue, probably about a 10-second long window. This window is determined by the beta decay following the fission and would be largely distance independent. As long as the background rate is sufficiently low, an antineutrino signal in that window would be a statistically credible indicator that the suspect seismic event involved nuclear fission. The study quantified the size of water-based antineutrino detectors that could confirm the nuclear nature of an explosion with a given yield, from a given standoff. While other detector materials such as liquid scintillator are possible for antineutrino detection from sources with little stand-off (e.g. a nuclear reactor), Water-Cherenkov detectors are the focus of this study due to the large masses required for nuclear explosion sources and large stand-offs. Results, derived from a detailed signal simulation and realistic background estimates, appear in Figure 3.

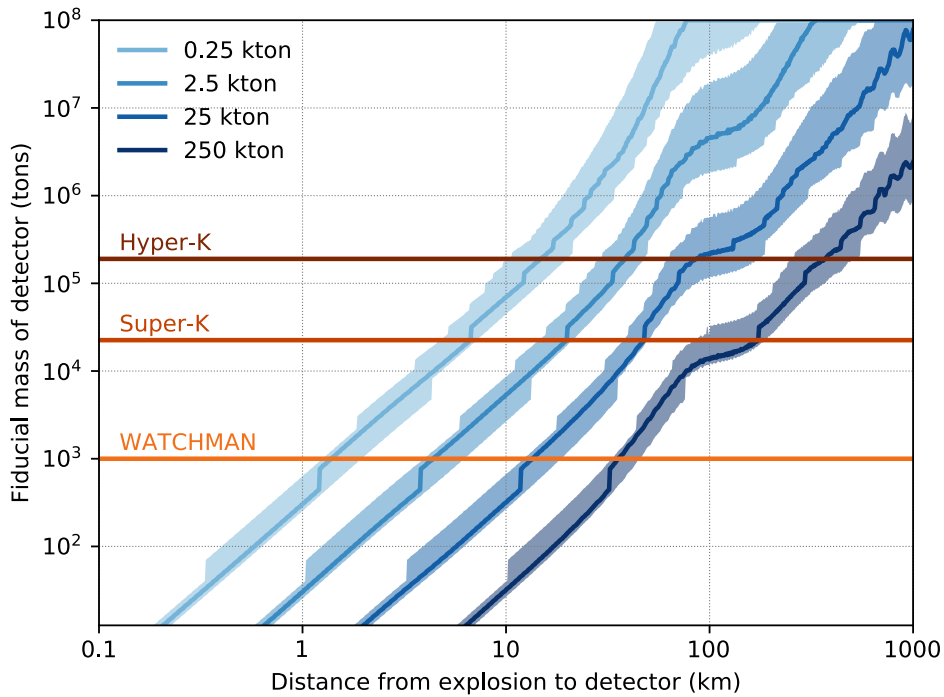

*Figure 3: Mass of a water-based neutrino detector that could give 90% probability of confirming fission yield for a suspect seismic event at 99% confidence level, for various true yields (blue curves). The step discontinuities come from the small number of discrete events required when backgrounds are low. The smooth waves come from neutrino flavor oscillations. The horizontal lines indicate the sizes of three planned or existing neutrino detectors: Hyper-Kamiokande (a very large detector planned for Japan), Super-Kamiokande (the largest existing liquid water-based neutrino detector, in Japan), and WATCHMAN (a detector planned for a nuclear reactor monitoring demonstration in the UK). Details of this figure appear in Carr et al., 2018.*

Even if we consider a scenario comparable to more recent nuclear tests of high-yield fission explosions, very large antineutrino detectors would be needed to confirm the nuclear nature, even from modest distances. As a baseline for a larger yield scenario, we consider a DPRK nuclear test with fission yield of 20 kT based on the 2016 DPRK tests (Pasyanos, 2018) and a 2.6 MTonne detector costing $5B (10 times the size of Hyper K (Hyper-Kamiokande, 2018) and 50 times the size of Super K) located at a distance of 500 km, roughly half way between Japan and North Korea (deployed sufficiently deep, such that the water shielding is enough to reduce the background from false-antineutrino events). Based on this scenario, we calculate that 4.6 events (on par with the level of events required to distinguish the antineutrino signals from background) would have been observed from the nuclear test in a 2.6 MTonne detector.

However, it is notable that nuclear testing at the 20-kT level would produce large seismic signals at 100 km and if radionuclides are released, large concentrations of radionuclides. An improvement to the global monitoring network at a distance of 100 km would require a very dense grid of antineutrino detectors compared to the approximate 1000-km grid size used in the IMS. Costs for such a system could exceed $1000B. Such an investment in infrastructure should rightfully be compared to the improvement that could be achieved by increasing the current radionuclide detection systems, the current means used to confirm

the nuclear nature of an explosive event, using the same amount of resources. Approximately $10^6$ radionuclide detection systems could be purchased for the equivalent cost. At this scale of investment, one current generation radionuclide sensor could be placed per 100 km$^2$, noting Manhattan Island is approximately 60 km$^2$ in area. This density is much smaller than the localization utilized by the CTBTO, which has a maximum inspection area under the CTBTO On-Site Inspection regime of 1000 km$^2$. For further comparison, the current IMS radionuclide sensor density is approximately one sensor per 100,000 km$^2$. At a density of sensitive radionuclide systems of 1000 times the current density, detection thresholds using radionuclide systems alone could prove sufficient for detection, considering such an array would have a sensor on average within 10 km from the test site.

*Use Case #1b: Augmenting a Global Monitoring Network*

If we require that the network of detectors be on par with current IMS nuclear explosion monitoring capabilities (National Academy of Sciences, 2002), we can assess the scenario of a ~1 kT explosion that is detonated at an unknown location throughout the world and is not detected by the current IMS. In this scenario, a monitoring distance can be obtained from the radionuclide spacing in the IMS. The spacing between radionuclide stations is approximately 2000 km, so we use a monitoring distance of 1000 km. We calculate the antineutrino flux at the detector based on a source of antineutrinos above the inverse beta decay from a nuclear explosion ($7.25 \times 10^{22}$ antineutrinos/1 kT):

*Equation 1: Calculation of the antineutrino ($\bar{\nu}$) flux at the detector for global monitoring*

$$\text{Antineutrino Flux at Detector} = \frac{\text{Antineutrino } (\bar{\nu}) \text{ Source}}{(\text{Distance from explosion})^2} = \frac{7.25 \times 10^{22} \bar{\nu}}{(1000 \text{ km})^2} = 7250000 \frac{\bar{\nu}}{cm^2}$$

For inverse beta decay, once above the threshold of 1.802, the average cross section of interaction is approximately $10^{-44}$ cm$^2$ (with the energy averaged cross section of $5.92 \pm 0.14 \times 10^{-43}$ cm$^2$/fission from a reactor). Using the interaction cross section above the threshold and the density of the detector material (water), we calculate the number of antineutrino events per gram of detector mass:

*Equation 2: Calculation of the antineutrino interaction rate within the detector medium (water) based on the number of protons per gram of water.*

$$\text{Antineutrino interaction rate} = 7250000 \frac{\bar{\nu}}{cm^2} \cdot 10^{-44} cm^2 \cdot 3.3 \times 10^{23} \frac{1}{g} = 2.4 \times 10^{-14} \frac{\bar{\nu} \text{ events}}{g}$$

Based on the background rates for an antineutrino detector, it is suggested that a burst of 5 events is required to statistically correlate the antineutrinos to the same source (Bernstein, 2001). The detector size is then calculated as:

*Equation 3: Calculation of the required antineutrino detector size for the required event burst observed within the detector for a global monitoring network.*

$$\text{Detector size} = \frac{\text{Event burst size}}{\bar{\nu} \text{ interaction rate}} = \frac{5 \text{ events}}{2.4 \times 10^{-14} \frac{\bar{\nu} \text{ events}}{\text{g}}} = 2.06 \times 10^{14} \text{g} = 200 \text{ MTonnes} = 200,000,000 \ m^3$$

At a detector size of 200 MTonnes, this would be the equivalent of 200 Empire State Building size detectors. Based on previous large-scale antineutrino detectors, a cost of ~$100M for a 50 kTonne is assumed, meaning that a single antineutrino detector for the global monitoring of nuclear explosions would cost ~$400B. At a value of $400B apiece, the cost would be 400 times the cost of the entire 1-billion-dollar IMS network. An extremely well contained underground explosion might still be below the detection limits of a higher density IMS, but the overall cost-to-benefit ratio seems much more favorable than for an antineutrino detector network.

Recall that both radionuclide and antineutrino detectors could be used to verify if an explosion was nuclear in nature. While the IMS has both particulate and radioxenon radionuclide detectors, the focus here is on the radioxenon as it is more likely to be released from an underground nuclear test. There are currently 40 locations for radioxenon detectors, with the potential for this value to be increased to 80 sensors, greatly increasing the sensitivity for determining if an explosion was nuclear in nature. As a price comparison, we look at the cost of the radioxenon detection systems. It is known that the radioxenon detectors cost ~$1M per system, so it would cost approximately $40M to double the density of radioxenon detectors throughout the network.

Due to the complexity and extent of the infrastructure required for an antineutrino detector compared to that of the rest of the IMS, one could expect that State parties to the CTBT may be more hesitant to allow an antineutrino detector in their country compared to the IMS stations. From this, it is clear that the locations of land-based antineutrino detectors would likely be limited, and an alternative of water-based locations would need to be investigated (Cicenas, 2011). The water-based locations would put a larger distance between the detector and the closest crustal source of background antineutrinos. However, the same requirements on size and detector depth would need to be followed for a water-based location, with the detector construction (vessel) and detection medium still being the same as on land. Therefore, water-borne locations may expand the list of potential locations, but it doesn't solve the issues of distance from the antineutrino source (nuclear explosion) for mid-continent nuclear test sites or the cost associated with developing and deploying such a detector.

Meeting this scenario would require that the detectors be spaced approximately 2000 km apart on the coast. It should be noted, that there are scenarios where the detector will be farther than 1000 km from a potential explosion site if only coastal detector locations are utilized, resulting in approximately 100 antineutrino detectors. With each detector costing 400 billion dollars, the total network cost would be approximately 40 trillion dollars, far beyond a feasible cost for a detector network (antineutrino or non-antineutrino).

Due to the detector size and cost required for antineutrino detection, it is not practical to think a large network of detectors would be deployed around the world. For this reason, the detectors would need to be placed at strategic locations in an effort to detect illicit nuclear explosions, although they would still be more costly and potentially less effective than the alternatives (such as an expanded IMS).

**Use Case #2: Test-Site Transparency**

The second use case of interest is the scenario in which an antineutrino detector is deployed on a nuclear test-site as part of a test-site transparency agreement. This case alleviates the concern of the first use case because only a few detectors are required worldwide and they are deployed at known test locations with a specific monitoring objective. In this situation, if a 1-kT explosion were to take place, it would need to be detected within the confines of the declared nuclear test site. Based on the dimensions of the former Nevada Nuclear Test Site, a detection stand-off can be estimated to be on the order of 50 km. One key difference is the likelihood that the backgrounds are higher due to the detector being located mid-continent instead of underwater. The antineutrino backgrounds would increase by approximately a factor of 3 (with hot spots in certain areas). While the detection criteria would increase depending on increased backgrounds of the location (Currie, 1968), we assume that the same detection criteria can be utilized and require a 5-neutrino-event signal for detection:

*Equation 4: Calculation of the antineutrino ($\bar{\nu}$) flux at the detector for test-site transparency*

$$\text{Antineutrino Flux at Detector} = \frac{\text{Antineutrino }(\bar{\nu})\text{ Source}}{(\text{Distance from explosion})^2} = \frac{7.25 \times 10^{22} \bar{\nu}}{(50 \text{ km})^2} = 2.9 \times 10^9 \frac{\bar{\nu}}{cm^2}$$

*Equation 5: Calculation of the antineutrino interaction rate within the detector medium (water) based on the number of protons per gram of water.*

$$\text{Antineutrino interaction rate} = 2.9 \times 10^9 \frac{\bar{\nu}}{cm^2} \cdot 10^{-44} \text{ cm}^2 \cdot 3.3 \times 10^{23} \frac{1}{g} = 9.7 \times 10^{-12} \frac{\bar{\nu} \text{ events}}{g}$$

*Equation 6: Calculation of the required antineutrino detector size for the required event burst observed within the detector for test-site transparency.*

$$\text{Detector size} = \frac{\text{Event burst size}}{\bar{\nu} \text{ interaction rate}} = \frac{5 \text{ events}}{9.7 \times 10^{-12} \frac{\bar{\nu} \text{ events}}{g}} = 500 \text{ kTonnes} = 500,000 \text{ } m^3$$

For reference, this detector would be ~twice the size of Hyper-K, ~200 times the size of the Statue of Liberty and cost 1 billion dollars. In the instance of being able to have additional technologies available to correlate the events (such as seismic cueing), the detection threshold could drop from 5-neutrino-events to 2-neutrino events. In this instance, the detector would be dropped to approximately the size of Hyper-K and 400 million dollars.

An alternative scenario would be the implementation of current monitoring technologies for test-site transparency, with radioxenon detectors being the most comparable for verification of the nuclear nature of an explosion. If we assume that 0.1% of all of the radioxenon produced from a 1 kT nuclear explosion was released (estimated from possible scenarios (Ringbom, 2014)), then we can calculate the detection threshold for a radioxenon detector.

*Equation 7: Calculation of the radioxenon detector system sensitivity at a distance of 50 km for test-site transparency. The calculation is based on the total amount of radioxenon produced from a 1 kT nuclear explosion along with an assumed release fraction and an average dilution value at 50 km.*

$$1 \text{ kT} \sim 10^{15} \text{ Bq of } {}^{133}\text{Xe} \cdot 10^{-3} \text{ release fraction} = 10^{12} \text{ Bq of } {}^{133}\text{Xe} \cdot \text{dilution of } 10^{10} \frac{1}{m^3} = 10^2 \frac{\text{Bq}}{\text{m}^3}$$

Based on the current detection threshold of 0.15 mBq/m$^3$ for the Xenon International system (LePetit, 2013), a radioxenon detector placed near the antineutrino detector would have a detection threshold of ~$1.5 \times 10^{-6}$ kT. This improved detection threshold suggests that a xenon detector would have better sensitivity than an antineutrino detector monitoring the same test-site area. Alternatively, this suggests that a radioxenon detector for test-site transparency could be sensitive to a release fraction of just $1.5 \times 10^{-9}$ of a 1 kT nuclear explosion.

**Use Case #3: Applications Beyond Detection**

Other hypothetical roles for antineutrinos are in providing information about a fission explosion's yield. Antineutrinos are a theoretically attractive approach for estimating fission yield, because the total antineutrino emission is basically proportional to fission yield. In contrast, the magnitude of seismic signals depends heavily on the depth of an underground explosion and other geological factors. However, making an antineutrino-based estimate of fission yield that is precision competitive with seismic estimates requires a stronger signal than confirming the presence of fission. This application would thus require detectors larger and more expensive than those depicted in Figure 3, (Carr et al., 2018).

In theory, a discrepancy between antineutrino-based and seismic-based yield estimates could indicate the presence of some fusion yield (if the seismic-based, total-energy yield is known to be larger than the antineutrino-based, fission-only yield) or intentional masking of the seismic signal by underground cavity engineering (if the seismic signal is significantly smaller than the antineutrino signal). The stringent requirements on both the antineutrino and seismic signals make this application even less conceivable than antineutrino-based yield constraints.

**IV. Conclusions**

While there are scenarios where antineutrino detectors are capable of monitoring for nuclear explosions, these scenarios appear to be limited unless budgets are nearly unlimited, or some other extremely special case is identified in which costs can be leveraged by other missions such as basic science. Another important point that one must remember when considering the use of antineutrino detectors is the improvements that could be obtained by expanding the IMS using current technologies with the cost a fraction of the cost that would be needed to deploy a limited number of antineutrino detectors. While a fully contained nuclear explosion may evade radionuclide detection of a higher density IMS system, the risk of this is much smaller than the cost of incorporating the slight potential benefit of an antineutrino network.

To obtain broad detection of nuclear explosions, the size and cost (200 MTonne and $400B) is far beyond what is practical. In the instance of test-site transparency, the size and cost (500 kTonne and $1B) and the time to build a detector limit the applicability. As the fundamental physics detectors continue to grow, there may be instances where they could detect a large nuclear explosion, but these instances will be rare and should not be considered to be part of routine nuclear explosion monitoring.

**Acknowledgments**

The authors acknowledge the support of the National Nuclear Security Administration Office of Defense Nuclear Nonproliferation Research and Development, U.S. Department of Energy, for funding this work. Any subjective views or opinions expressed in the paper do not necessarily represent the views of the U.S. Department of Energy or the United States Government.